\documentstyle[preprint,prd,aps,rotating]{revtex}
\draft
\begin{document}
\tightenlines
\input epsf
\title{\begin{flushright} {\rm\small HUB-EP-97/90}\end{flushright}
Mass spectrum of orbitally and radially excited heavy-light
mesons in the relativistic quark model}
\renewcommand{\thefootnote}{\fnsymbol{footnote}}
\author{ D. Ebert
~and V. O. Galkin\thanks{On leave of absence 
from Russian Academy of Sciences,
Scientific Council for Cybernetics,
Vavilov Street 40, Moscow 117333, Russia.}}

\address{Institut f\"ur Physik, Humboldt--Universit\"at zu Berlin,
Invalidenstr.110, D-10115 Berlin, Germany}

\author{R. N. Faustov}
\address{Russian Academy of Sciences, Scientific Council for
Cybernetics, Vavilov Street 40, Moscow 117333, Russia}

\date{\today}
\maketitle
\begin{abstract}
The mass spectrum of orbitally and radially excited states of
 $B$ and $D$ mesons is calculated in the
framework of the relativistic quark model.  The
expansion in inverse powers of the heavy quark mass is
carried out up to the first
order, while the light quark is treated without expansion. We 
find that the relativistic treatment of the light quark plays an
important role. Different patterns of $P$ level inversion are
discussed.  The obtained masses of orbitally and radially
excited states are in accord with available experimental data and 
heavy quark symmetry relations.
\end{abstract}


\section{Introduction}  
The investigation of the properties of excited heavy-light mesons
represents an interesting and  important problem.
The experimental data on the orbitally 
\cite{orb} and radially \cite{rad} excited $B$ and $D$ meson states are
becoming available now. The $B$ and $c-\tau$ factories will provide more
accurate and comprehensive data for these states. On the other hand,
the presence of the heavy quark in these systems considerably
simplifies the theoretical description of heavy-light mesons.
The heavy quark symmetry \cite{iw}, arising in the limit of infinitely
heavy quark mass $m_Q$, imposes strict
constraints on the mass spectrum of heavy-light mesons \cite{iw2}. In this
limit the heavy quark mass and spin decouples and all the meson properties 
are determined by light degrees of freedom alone. As a result the heavy quark
spin degeneracy of the levels emerges. The light quark's spin couples 
with its orbital momentum, resulting  for $S$-wave mesons in two 
degenerate  $j=1/2$ states, corresponding to $1^-$ and $0^-$ \footnote{
We use the $J^P$ notation, where $J=j\pm 1/2$ and $P$ are the total 
angular momentum and parity of the meson.}. For 
$P$-wave mesons two degenerate  $j=3/2$ states, $(1^+,2^+)$, and 
two degenerate $j=1/2$ states, $(0^+,1^+)$, arise. The heavy quark
symmetry constrains also the strong decays of these states. The
decay rates of the $P$-wave states in degenerate doublets should be 
the same. The states with $j=1/2$ are expected to be wide, because
they decay in $S$-wave, while $j=3/2$ mesons are narrow, because 
they decay in $D$-wave.  Since the
real $b$ and $c$ quarks are not infinitely heavy, the corrections
in inverse powers of the heavy quark mass turn out to be important.
These corrections break down the degeneracy of the levels. The 
heavy quark effective theory (HQET) (see e.g. \cite{n} 
and references therein) provides a systematic
method for treating $1/m_Q$ corrections. However, in order to obtain 
quantitative predictions it is necessary to combine it with 
some dynamical nonperturbative approaches.  

Many different approaches have been used for the calculation of
orbital and radial excitations of heavy-light mesons \cite{gi,ehq,i,thq}. 
However,
almost in all of them the expansion in inverse powers not only of
the heavy quark mass ($m_Q$) but also  in inverse powers of the 
light quark mass  ($m_q$) is carried out. 
The estimates of the light quark velocity in these mesons show that
the light quark is highly relativistic ($v/c\sim 0.7\div 0.8$). Thus
the nonrelativistic approximation is not adequate for the light quark 
and one cannot guarantee the numerical accuracy of the expansion in inverse
powers of the light quark mass. In this paper we use the relativistic 
quark model \cite{mass,fgm,efg} for the
calculation of the masses of orbitally and radially excited $B$ and
$D$ mesons without employing the expansion in  $1/m_q$. Thus the light
quark is treated fully relativistically. Concerning the heavy quark 
we apply the 
expansion in $1/m_Q$ up to the first order.
Our relativistic quark model is based on the quasipotential approach
in quantum field theory. It has been used for the calculation of
heavy quarkonia mass spectrum \cite{mass} and electroweak decays of
heavy mesons \cite{fgm,efg}. Recent nonperturbative investigations
indicate that the confining potential cannot be simply  scalar. This
is in agreement with our model assumptions on the Lorentz structure
of quark confinement. The comparison of our model results with model
independent constraints of HQET is given in \cite{fg}. 
      
The paper is organized as follows. In Sec.~\ref{rqm} we describe 
our relativistic quark model giving special emphasize on the role 
of the Lorentz-structure of quark confinement. In Sec.~\ref{qp} we construct
the quasipotential of the interaction of a light quark with a heavy antiquark. 
We use the heavy quark $1/m_Q$ expansion to simplify the construction. The
light quark is treated relativistically. First we consider the $m_Q\to
\infty$ limit and then the corrections of the first order in $1/m_Q$.
We present the predictions of our model for orbitally and radially 
excited states of $D$, $D_s$, $B$ and $B_s$
mesons. In Sec.~\ref{rd} we compare our results with the heavy quark
symmetry and other quark model predictions as well as available 
experimental data. Section~\ref{concl} contains our conclusions.

\section{Relativistic quark model}  
\label{rqm}

In the quasipotential approach a meson is described by the wave
function of the bound quark-antiquark state, which satisfies the
quasipotential equation \cite{3} of the Schr\"odinger type \cite{4}
\begin{equation}
\label{quas}
{\left(\frac{b^2(M)}{2\mu_{R}}-\frac{{\bf
p}^2}{2\mu_{R}}\right)\Psi_{M}({\bf p})} =\int\frac{d^3 q}{(2\pi)^3}
 V({\bf p,q};M)\Psi_{M}({\bf q}),
\end{equation}
where the relativistic reduced mass is
\begin{equation}
\mu_{R}=\frac{E_qE_Q}{E_q+E_Q}=\frac{M^4-(m^2_q-m^2_Q)^2}{4M^3},
\end{equation}
and $E_q$, $E_Q$ are given by
\begin{equation}
\label{ee}
E_q=\frac{M^2-m_Q^2+m_q^2}{2M}, \quad E_Q=\frac{M^2-m_q^2+m_Q^2}{2M},
\end{equation}
here $M=E_q+E_Q$ is the meson mass, $m_{q,Q}$ are the masses of light
and heavy quarks, and ${\bf p}$ is their relative momentum.  
In the centre of mass system the relative momentum squared on mass shell 
reads
\begin{equation}
{b^2(M) }
=\frac{[M^2-(m_q+m_Q)^2][M^2-(m_q-m_Q)^2]}{4M^2}.
\end{equation}

The kernel 
$V({\bf p,q};M)$ in Eq.~(\ref{quas}) is the quasipotential operator of
the quark-antiquark interaction. It is constructed with the help of the
off-mass-shell scattering amplitude, projected onto the positive
energy states. An important role in this construction is played 
by the Lorentz-structure of the confining quark-antiquark interaction
in the meson. In the literature there is no consent on this item. For a 
long time the scalar confining kernel has been considered to be the
most appropriate one \cite{scal}. The main argument in favour of this
choice is based on the nature of the heavy quark spin-orbit potential.
The scalar potential gives a vanishing long-range magnetic 
contribution, which is in agreement with the picture
of quark confinement of Ref.~\cite{buch}, and allows to get the fine
structure for heavy quarkonia in accord with experimental data. 
However, the calculations of electroweak decay rates of heavy mesons 
with a scalar confining potential alone yield results which are in  worse 
agreement with data than for a vector potential \cite{mb,gf}. 
The  radiative
$M1$-transitions in quarkonia such as e.~g. $J/\psi\to \eta_c
\gamma$ are the most sensitive
for the Lorentz-structure of the confining potential. 
The relativistic corrections for 
these decays arising from vector and scalar potentials have different
signs \cite{mb,gf}. In particular, as it has been 
shown in Ref.~\cite{gf}, agreement
with experiments for these decays can be achieved only for a mixture
of vector and scalar potentials. In this context, it is worth remarking,
that the recent study of the $q\bar q$ 
interaction in the Wilson loop approach \cite{bv} indicates that
it cannot be considered as simply a scalar. Moreover, the found
structure of spin-independent relativistic corrections is not 
compatible with a scalar potential. A similar conclusion
has been obtained in Ref.~\cite{ss} on the basis of a Foldy-Wouthuysen 
reduction of the full Coulomb gauge Hamiltonian of QCD. There, the 
Lorentz-structure of confinement has been found to be of vector
nature. The scalar character of spin splittings in heavy quarkonia
in this approach is dynamically generated through the interaction
with collective gluonic degrees of freedom. 

All these new results are in agreement with the assumptions of the
relativistic quark model approach \cite{gf,mass}. Indeed, in 
constructing the quasipotential of quark-antiquark interaction 
we have assumed that the effective
interaction is the sum of the usual one-gluon exchange term with the mixture
of long-range vector and scalar linear confining potentials, where
the vector confining potential
contains the Pauli interaction. The quasipotential is then defined by
\cite{mass}  
\begin{eqnarray}
\label{qpot}
V({\bf p,q};M)&=&\bar{u}_q(p)
\bar{u}_Q(-p)\Bigg\{\frac{4}{3}\alpha_SD_{ \mu\nu}({\bf
k})\gamma_q^{\mu}\gamma_Q^{\nu}\cr
& & +V^V_{\rm conf}({\bf k})\Gamma_q^{\mu}
\Gamma_{Q;\mu}+V^S_{\rm conf}({\bf
k})\Bigg\}u_q(q)u_Q(-q),
\end{eqnarray}
where $\alpha_S$ is the QCD coupling constant, $D_{\mu\nu}$ is the
gluon propagator in the Coulomb gauge
\begin{equation}
D^{00}({\bf k})=-\frac{4\pi}{{\bf k}^2}, \quad D^{ij}({\bf k})=
-\frac{4\pi}{k^2}\left(\delta^{ij}-\frac{k^ik^j}{{\bf k}^2}\right),
\quad D^{0i}=D^{i0}=0,
\end{equation}
and ${\bf k=p-q}$; $\gamma_{\mu}$ and $u(p)$ are 
the Dirac matrices and spinors
\begin{equation}
\label{spinor}
u^\lambda({p})=\sqrt{\frac{\epsilon(p)+m}{2\epsilon(p)}}
{1\choose \frac{\bbox{\sigma p}}{\epsilon(p)+m}}\chi^\lambda,
\end{equation}
with $\epsilon(p)=\sqrt{p^2+m^2}$.
The effective long-range vector vertex is
given by
\begin{equation}
\Gamma_{\mu}({\bf k})=\gamma_{\mu}+
\frac{i\kappa}{2m}\sigma_{\mu\nu}k^{\nu},
\end{equation}
where $\kappa$ is the Pauli interaction constant characterizing the
anomalous chromomagnetic moment of quarks. Vector and
scalar confining potentials in the nonrelativistic limit reduce to
\begin{eqnarray}
V^V_{\rm conf}(r)&=&(1-\varepsilon)(Ar+B),\nonumber\\
V^S_{\rm conf}(r)& =&\varepsilon(Ar+B),
\end{eqnarray}
reproducing 
\begin{equation}
V_{\rm conf}(r)=V^S_{\rm conf}(r)+
V^V_{\rm conf}(r)=Ar+B,
\end{equation}
where $\varepsilon$ is the mixing coefficient. 

The expression for the quasipotential for the heavy quarkonia,
expanded in $v^2/c^2$, can be found in Ref.~\cite{mass}. The 
structure of the spin-dependent interaction is in an agreement with
the parameterization of Eichten and Feinberg \cite{ef}.  
All the parameters of
our model like quark masses, parameters of linear confining potential
$A$ and $B$, mixing coefficient $\varepsilon$ and anomalous
chromomagnetic quark moment $\kappa$ were fixed from the analysis of
heavy quarkonia masses \cite{mass} and radiative decays \cite{gf}. 
The quark masses
$m_b=4.88$ GeV, $m_c=1.55$ GeV, $m_s=0.50$ GeV, $m_{u,d}=0.33$ GeV and
the parameters of the linear potential $A=0.18$ GeV$^2$ and $B=-0.30$ GeV
have standard values of quark models.  The value of the mixing
coefficient of vector and scalar confining potentials $\varepsilon=-1$
has been determined from the consideration of the heavy quark expansion
\cite{fg} and meson radiative decays \cite{gf}.
Finally, the universal Pauli interaction constant $\kappa=-1$ has been
fixed from the analysis of the fine splitting of heavy quarkonia ${
}^3P_J$- states \cite{mass}. Note that the 
long-range  magnetic contribution to the potential in our model
is proportional to $(1+\kappa)$ and thus vanishes for the 
chosen value of $\kappa=-1$.
\section{Quasipotential of the interaction of a light quark  with a heavy
antiquark} \label{qp}
The expression for the quasipotential (\ref{qpot}) can, in principle, be
used for arbitrary quark masses.  The substitution
of the Dirac spinors (\ref{spinor}) into (\ref{qpot}) results in an extremely
nonlocal potential in the configuration space. Clearly, it is very hard to 
deal with such potentials without any simplifying expansion.
Fortunately, in the 
case of heavy-light mesons, one can carry out (following HQET)
the expansion in inverse 
powers of the heavy quark mass $m_Q$. The leading terms then follow
in the limit $m_Q\to \infty$. 
\subsection{Infinitely heavy quark limit}
In the limit $m_Q\to\infty$ the heavy quark Dirac spinor
has only an upper component. Thus only the zeroth component of
Dirac matrices in (\ref{qpot}) contributes, and we 
get
\begin{eqnarray}
\label{ipot}
V({\bf p},{\bf q};M)&=& \bar u_q(p)\Bigg\{-\frac43\alpha_s
\frac{4\pi}{{\bf k}^2}\gamma_q^0\cr
& & +V_{\rm conf}^V({\bf k})\left[\gamma_q^0+\frac{\kappa}{2m_q}
\gamma_q^0(\bbox{\gamma k})\right]+V_{\rm conf}^S({\bf k})
\Bigg\}u_q(q).
\end{eqnarray}
The resulting quasipotential is still nonlocal in configuration
space. However, taking
into account that $B$ and $D$ mesons are weakly bound~\footnote{The 
sum of constituent quark masses $m_Q+m_q$ is very close to the ground
state meson mass $M$.}, we can replace $\epsilon_q(p) \to E_q$
in  the Dirac spinors (\ref{spinor}). 
Such simplifying substitution is widely used
in quantum electrodynamics \cite{bs,mf,km} and introduces only minor 
corrections of order of the ratio of the binding energy $\langle V
\rangle$ to $E_q$. This substitution makes the Fourier 
transformation of the potential (\ref{ipot}) local and
works well for the confining part of the potential. However, it leads to 
a fictitious singularity $1/r^3$ at the origin arising from the 
one-gluon exchange part, which is absent in the initial potential.
Note that this singularity is not important if the perturbation
theory in $1/m_q$ is used. As we are not going to expand our
potential in $1/m_q$, additional analysis is necessary. The explicit 
one-gluon contribution to (\ref{ipot}) is given by
\begin{eqnarray}
\label{sing}
\bar u_q(p)\left(-\frac43\alpha_s\frac{4\pi}{{\bf k}^2}\gamma_q^0\right)u_q(q)
&=&\left(-\frac43\alpha_s\frac{4\pi}{{\bf k}^2}\right)
\sqrt{\frac{\epsilon_q(p)+m_q}{2\epsilon_q(p)}}
\sqrt{\frac{\epsilon_q(q)
+m_q}{2\epsilon_q(q)}}\cr
& &\times\Bigg(1 +
\frac{{\bf pq}+i\bbox{\sigma}_q[{\bf p q}]}
{[\epsilon_q(p)+m_q][\epsilon_q(q)+m_q]}\Bigg).
\end{eqnarray}    
The most interesting is the last term in this expression, which
comes from the lower components of the spinors $u_q$. In configuration
space, it has an $1/r$ asymptotics for $r\to 0$ and decreases
as $1/r^3$ for $r\to \infty$. If we simply replace $\epsilon_q
\to E_q$, then we get the asymptotics $1/r^3$, which is
correct  at infinity but too singular at the origin. To cure
this asymptotics, let us notice that if binding effects 
are taken into account, it is necessary to replace $\epsilon_q
\to E_q-V$, where $V$ is the quark interaction potential. If we 
consider larger distances of order of the hadron radius, the contributions
of $V$ are of order of the (small) binding energy and can be neglected. But
for $r\to 0$ the Coulomb singularity in $V$ becomes important. Thus,
we keep  this contribution (up to the  first order
in the binding energy) in the last term in (\ref{sing}). As a result
we find that the Coulomb-like one-gluon potential $V_{\rm Coul}(r)=
-\frac43\frac{\alpha_s}{r}$ in this term
should be replaced by
\begin{equation}
\tilde  V_{\rm Coul}(r)=V_{\rm Coul}(r)\frac{1}{\left(1+\frac43\frac{
\alpha_s}{E_q}\frac{1}{r}\right)\left(1+\frac43\frac{\alpha_s}{
E_q+m_q}\frac{1}{r}\right)},
\end{equation}        
leading to the correct asymptotics both at the origin and infinity.

The resulting local quark-antiquark potential for $m_Q\to \infty$
can be presented in configuration space in the following form
\begin{eqnarray}
\label{vinf}
V_{m_Q\to \infty}(r)&=& \frac{E_q+m_q}{2E_q}\Bigg[V_{\rm Coul}(r)
+V_{\rm conf}(r) + \frac{1}{(E_q+m_q)^2}\Bigg\{{\bf p}[\tilde 
V_{\rm Coul}(r)\cr
& & +V_{\rm conf}^V(r)-V_{\rm conf}^S(r)]{\bf p}
-\frac{E_q+m_q}{2m_q}\Delta V_{\rm
conf}^V(r)[1-(1+\kappa)] \cr
& & +\frac{2}{r}\left(\tilde V_{\rm Coul}'(r)-V_{\rm conf}'^S(r) -
V_{\rm conf}'^V(r)\left[\frac{E_q}{m_q}
-2(1+\kappa)\frac{E_q+m_q}{2m_q}\right]\right)
{\bf L}{\bf S}_q\Bigg\}\Bigg].
\end{eqnarray}
Here the prime denotes differentiation with respect to $r$, ${\bf L}$ is
the orbital momentum, 
and ${\bf S}_q$ is the spin operator of the light quark. Note that
the last term in (\ref{vinf}) is of the same order as the first two
terms and thus cannot be treated perturbatively.

In the infinitely heavy quark limit the quasipotential equation
(\ref{quas}) in configuration space becomes
\begin{equation}
\label{iquas}
\left(\frac{E_q^2-m_q^2}{2E_q}-\frac{{\bf p}^2}{2E_q}\right)
\Psi_M(r)=V_{m_Q\to\infty}(r)\Psi_M(r),
\end{equation}
and the mass of the meson is given by $M=m_Q+E_q$.

Solving (\ref{iquas}) numerically \cite{fgtv} we 
get the eigenvalues $E_q$ 
and the wave functions $\Psi_M$. The obtained results are presented
in Table~\ref{eq}.
 We see that the heavy
quark spin decouples in  the limit $m_Q\to\infty$, and thus
we get the number of degenerated states in accord with the
heavy quark symmetry prediction.  
  
\subsection{$1/m_Q$ corrections}
The heavy quark symmetry degeneracy of states is broken by $1/m_Q$
corrections. The corrections of order $1/m_Q$ to the potential
(\ref{vinf}) arise from the lower components of heavy quark Dirac 
spinors in (\ref{qpot}). The matrix elements of the spatial components 
of Dirac matrices are not zero now. Thus the contributions at first
order in $1/m_Q$ come from the one-gluon-exchange potential and the
vector confining potential, while the scalar potential gives no 
contribution at first order. The resulting $1/m_Q$ correction to  
the heavy quark potential (\ref{vinf}) is given by the following
expression
\begin{eqnarray}
\label{vcor}
\delta V_{1/m_Q}(r)&=&\frac{1}{E_qm_Q}\Bigg\{{\bf p}\left[V_{\rm
Coul}(r)+V^V_{\rm conf}(r)\right]{\bf p}
+V'_{\rm Coul}(r)\frac{{\bf L}^2}
{2r}\cr
&&-\frac{1}{4}\Delta V^V_{\rm conf}(r)+\left[\frac{1}{r}V'_{\rm 
Coul}(r)+\frac{(1+\kappa)}{r}V'^V_{\rm conf}(r)\right]{\bf LS}\cr
& & +\frac13\bigglb(\frac{1}{r}V'_{\rm Coul}(r)-V''_{\rm Coul}(r)
+(1+\kappa)^2\left[\frac{1}{r}V'^V_{\rm conf}(r)-V''^V_{\rm conf}(r)
\right]\biggrb)\cr
&&\times\left[-{\bf S}_q{\bf S}_Q+\frac{3}{r^2}({\bf S}_q{\bf
r})({\bf S}_Q{\bf r})\right]\cr
& &+\frac23\left[\Delta V_{\rm Coul}(r)+(1+\kappa)^2\Delta V^V_{\rm 
conf}(r)\right]{\bf S}_Q{\bf S}_q\Bigg\},
\end{eqnarray}
where ${\bf S}={\bf S}_q+{\bf S}_Q$ is the total spin. The first three terms
in (\ref{vcor}) represent spin-independent corrections, the fourth
term is responsible for the spin-orbit interaction, the fifth one is 
the tensor interaction and the last one is the spin-spin interaction.
It is necessary to note that the confining vector interaction gives a
contribution to the spin-dependent part which is proportional to
$(1+\kappa)$ or $(1+\kappa)^2$. Thus it vanishes for the chosen value
of $\kappa=-1$, while the confining vector contribution to 
the spin-independent part is non zero. 

The quasipotential at $1/m_Q$ order is
given by the sum of $V_{m_Q\to \infty}(r)$
from (\ref{vinf}) and $\delta V_{1/m_Q}(r)$ from (\ref{vcor}). By 
substituting it in the quasipotential equation (\ref{quas}) and 
treating the $1/m_Q$ correction term $\delta V_{1/m_Q}(r)$ using
perturbation theory, we are now able to  calculate the
mass spectrum of $D$, $D_s$, $B$ and $B_s$ mesons with the account 
of $1/m_Q$ corrections. The results of our calculations are presented
in Tables~\ref{md}-\ref{mbs}.

\section{Results and discussion}  \label{rd} 
Let us compare the obtained results with model independent
predictions of HQET. In HQET 
the heavy-light meson mass $M$ of ground-state pseudoscalar
and vector mesons is given by \cite{n}
\begin{equation}
\label{mhqet}
M=\mu_Q+\bar \Lambda-\frac{\lambda_1-2\left[J(J+1)-\frac32
\right]\lambda_2}{2\mu_Q} 
+O(1/\mu_Q^2),
\end{equation}
where the parameter $\bar \Lambda$ represents contributions arising from 
terms in the HQET Lagrangian which are independent of the heavy
quark mass. The terms $\lambda_1$ and $\lambda_2$ parametrize 
contributions from the kinetic energy and chromo-magnetic interaction.

Note that the HQET heavy quark mass $\mu_Q$ is in principle different from 
our mass $m_Q$ by the simple reason, that HQET uses the pole 
heavy quark mass, while quark models use the constituent quark mass.
Obviously, in the heavy quark symmetry limit this difference between  
constituent and pole heavy quark masses, however, disappears. In 
particular, we find the relation $E_q=\bar \Lambda+O(1/m_Q)$. Thus,
the values of 
$E_q$ listed in Table~\ref{eq} are just the values of the  HQET parameters
$\bar\Lambda$.

The structure of $1/m_Q$ corrections in our model is consistent
with (\ref{mhqet}). However, we find that the parameter corresponding
to $\lambda_1$ in our model contains not only the kinetic energy
part but also other terms originating from spin-independent
corrections in Eq.~(\ref{vcor})~\footnote{Note that, if we had used also an
expansion in inverse powers of the light quark mass, then, in
the static limit, we would get $\lambda_1=-\langle p^2\rangle$.}.
A similar result has been found in the relativistic quark model in
Ref.~\cite{sim}. The value of this parameter turns out to be
very sensitive to the value of the 
heavy quark mass. Indeed, assuming that the heavy quark pole mass $\mu_Q$ and
constituent quark mass $m_Q$ are related by \cite{sim}
\begin{equation}
\label{muq}
m_Q=\mu_Q -\frac{\lambda^Q}{2\mu_Q}+O(1/\mu_Q^2),
\end{equation}
we see that only $\lambda_1$ is influenced, and it is connected to
the quark model value $\lambda^{\rm QM}_1$ by
\begin{equation}
\label{lam1}
\lambda_1=\lambda^{\rm QM}_1+\lambda^Q.
\end{equation}
In our model, for constituent quark masses $m_b=4.88$ GeV and $m_c=1.55$ 
GeV, we find  $\lambda^{\rm QM}_1\approx 0.85$ GeV$^2$ for $B$ mesons and
$\lambda^{\rm QM}_1\approx 0.26$ GeV$^2$ for $D$ mesons. If we require the
parameter $\lambda_1$ to be equal to the mean value of $-\langle p^2
\rangle$ (in our model  $\langle p^2\rangle\approx 0.25$ GeV$^2$ for $B$ and
$D$ mesons), then from (\ref{muq}) and (\ref{lam1}) we get the heavy 
quark masses $\mu_b=4.75$ GeV and $\mu_c=1.40$ GeV. These values agree
with values of the heavy quark pole masses used in HQET.

The values of the parameter $\lambda_2$, which determines the hyperfine
splitting, coincide in HQET and quark models. We find $\lambda_2\approx0.112$ 
GeV$^2$ for $B$ mesons and $\lambda_2\approx 0.125$ GeV$^2$ for $D$ mesons. 

Heavy quark symmetry provides relations between excited states
of $B$ and $D$ mesons, such as
\begin{equation}
\label{rel1}
\bar M_{B_1}-\bar M_{D_1}=\bar M_{B_{s1}}-\bar M_{D_{s1}}=
\bar M_{B}-\bar M_{D}=\bar M_{B_s}-\bar M_{D_s}=m_b-m_c,
\end{equation}
where $\bar M_{B_1}=(3M_{B_1}+5M_{B_2})/8$, $\bar M_{B}=(M_B+3M_{B^*})/4$ 
are appropriate spin averaged $P$- and $S$-wave states. 
We get from Tables~\ref{md}-\ref{mbs} the following
values of mass splittings
\begin{eqnarray}
\bar M_{B_1}-\bar M_{D_1}&=&3.29\ {\rm GeV}\cr
\bar M_{B_{s1}}-\bar M_{D_{s1}}&=&3.30\ {\rm GeV}\cr
\bar M_{B}-\bar M_{D}&=&3.34\ {\rm GeV}\cr 
\bar M_{B_s}-\bar M_{D_s}&=&3.33\ {\rm GeV},
\end{eqnarray}
in agreement with (\ref{rel1}). There arise also the following
relations between hyperfine splittings of levels
\begin{equation}
\Delta M_{B}\equiv M_{B_2}-M_{B_1}=\frac{m_c}{m_b}\Delta
M_{D}\equiv \frac{m_c}{m_b}\left(M_{D_2}-M_{D_1}\right),
\end{equation}
and the same for $B_s$ and $D_s$ mesons as well as for $P_1-P_0$ states.
Our model predictions for these splittings are displayed in 
Table~\ref{splt}.

In Tables~\ref{md}-\ref{mbs} we compare our relativistic quark
model results for heavy-light meson masses with the 
predictions of other quark  models of Godfrey
and Isgur \cite{gi}, Isgur \cite{i},  Eichten, Hill and Quigg 
\cite{ehq} and experimental data \cite{PDG,orb,rad}. 
All these quark models  use the
expansion in inverse powers both of the heavy $m_Q$ and light $m_q$
quark masses for the $Q\bar q$ interaction potential. In Ref.~\cite{gi} 
some relativization of the potential has been put in by hand, such
as relativistic smearing of coordinates and replacing the factors
$1/m_q$ by $1/{\epsilon_q(p)}$. However, the resulting potential in
this approach accounts only for some of the relativistic effects, 
while the others, which are of the same order of magnitude, are missing.
The considerations of Refs.~\cite{i,ehq} are closely related. The 
heavy quark expansion is extended to light ($u,d,s$) quarks and the
experimental data on $P$ wave masses of $K$ mesons are used to obtain
predictions for $B$ and $D$ mesons.

In the paper~\cite{i} it is argued that the heavy quark spin $P$-wave
multiplets
with $j=1/2$ ($0^+,1^+$) and $j=3/2$ ($1^+,2^+$) in $B$ and $D$ mesons
are inverted \cite{hs}. 
The $2^+$ and $1^+$ states lie about 150 MeV below the 
$1^+$ and $0^+$ states.  In the limit $m_Q\to \infty$, we find 
the same inversion of these multiplets in our model, but the gap between
$j=1/2$ and $j=3/2$ states is smaller ($\sim 90$ MeV for $B$ and $D$ mesons
and $\sim 70$ MeV for $B_s$ and $D_s$ mesons), and  $1/m_Q$ corrections
reduce this gap further. However, the hyperfine splittings among the 
states in these multiplets turn out to be larger than in \cite{i}. 
As a result, the states from
the multiplets for $D$, $D_s$ and $B_s$ mesons
overlap in our model, however the heavy quark spin
averaged centres are still inverted (see Figs.~1-4). 
We obtain the following ordering of
$P$ states (with masses increasing from left to right): 
$B$ meson --- 
$1P_1(\frac32)$, $1P_2$, $1P_0$, $1P_1(\frac12)$;
$D_s$ meson --- $1P_0$, $1P_1(\frac32)$, $1P_2$, 
$1P_1(\frac12)$; $D$ and $B_s$
mesons --- $1P_1(\frac32)$, $1P_0$, $1P_2$, $1P_1(\frac12)$. Thus only for
$B$ meson we get the purely inverted pattern. Note that the model
\cite{gi} predicts the ordinary ordering of levels.

The results of our model agree well with available experimental data. The
experimental values in Tables~\ref{md}-\ref{mbs} for ground state 
and $P$-wave masses are taken from Refs.~\cite{PDG,orb}. For the radially
excited states we use the preliminary data from  DELPHI \cite{rad}.    
 
\section{Conclusions} \label{concl}
In this paper we have presented the calculation of the mass spectra
of orbitally and radially excited states
of heavy-light mesons in the framework of the relativistic quark model.
The main advantage of the proposed approach consists in the relativistic
treatment of the light quarks ($u,d,s$). We apply only the expansion in
inverse powers of the heavy quark ($b,c$) mass, which considerably
simplifies calculations. The infinitely heavy quark limit as well
as the first order $1/m_Q$ corrections are considered. Our model
respects the constraints imposed by heavy quark symmetry on the number
of levels and different splittings.       

We find that the heavy quark spin multiplets with $j=1/2$ ($0^+,1^+$)
and $j=3/2$ ($1^+,2^+$) are inverted in the $m_Q\to\infty$ limit.
This inversion is caused by the following reason: The confining potential
contribution to the spin-orbit term in (\ref{vinf}) exceeds the one-gluon
exchange contribution. Thus the sign before the spin-orbit term is negative,
and the level inversion emerges. However,
$1/m_Q$ corrections, which produce the hyperfine splittings of these
multiplets, are substantial. As a result the purely inverted pattern
of $P$ levels occurs only for the $B$ meson. 
For $D$ and $B_s$ mesons the levels
from these multiplets begin to overlap. This effect is more pronounced
in the $D_s$ meson, where the ordinary ordering is  restoring. 
Thus we see that the $D_s$ meson occupies an intermediate position
between heavy-light mesons and heavy-heavy mesons 
(quarkonia, $B_c$ meson). It is necessary to
note, that both the presence of the relativistic light quark and the
light-to-heavy quark mass ratio play an important role in the formation 
of level ordering patterns. The light quark determines the meson radius,
while the $m_q/m_Q$ ratio indicates the validity of the application
of the heavy quark symmetry limit. 

The found mass values of orbitally and radially excited heavy-light mesons
are in good  agreement with available experimental data. At present only
narrow $P$-wave $2^+,1^+$ ($j=3/2$) levels of $D$, $D_s$ and $B$ mesons 
have been measured. It will be very interesting to observe also $0^+,1^+$
($j=1/2$) levels, which is more complicated, because these states are 
expected to be broad. This will allow to determine the ordering of
$P$ levels and to test quark dynamics in a heavy-light meson. We plan
to apply the found meson wave functions for the calculation of 
semileptonic and nonleptonic decays of $B$ mesons into $P$-wave 
$D$ and $D_s$ mesons, which are important for the experimental
observation of these states.                

\acknowledgments
We are grateful to D.V. Antonov, A.B. Kaidalov, T. Lohse, V.I. Savrin, 
A.Yu. Simonov, and A.S. Vshivtsev for useful discussions
of the results.
One of the authors (V.O.G) gratefully acknowledges the warm hospitality
of the colleagues in the particle theory group of the Humboldt-University
extended to him during his stay there.
He was supported in part by {\it Deutsche
Forschungsgemeinschaft} under contract Eb 139/1-3 and
in part by {\it Russian Foundation for Fundamental Research}
 under Grant No.\ 96-02-17171.
R.N.F. was supported in part by {\it Russian Foundation for
Fundamental Research} under Grant No.\ 96-02-17171 and in part by
{\it Graduiertenkolleg Elementarteilchenphysik}.

\begin{table}
\caption{The values of $E_q$ and meson masses $M$ in the limit $m_Q\to\infty$
(in GeV). We use the notation $(n^jL)$ for meson states, where
$n$ is the radial quantum number, and $j$ is the total angular momentum of 
the light quark.}
\label{eq}

\begin{tabular}{ccccccccc}
 &\multicolumn{2}{c}{$D$}&\multicolumn{2}{c}{$D_s$}
&\multicolumn{2}{c}{$B$}&\multicolumn{2}{c}{$B_s$}\\
\hline
State& $E_{u,d}$ &$M$& $E_s$ &$M$& $E_{u,d}$&$M$& $E_s$ &$M$\\
\hline
$1^{\frac12}S$& 0.497 &2.047& 0.607 & 2.157 & 0.514 
& 5.394 & 0.628 & 5.508\\
$1^{\frac32}P$& 0.800 &2.350& 0.913 & 2.463 & 0.800
& 5.680 & 0.927 & 5.807\\
$1^{\frac12}P$& 0.886 &2.436& 0.985 & 2.535 & 0.898
& 5.778 & 0.998 & 5.878\\
$2^{\frac12}S$& 0.940 &2.490& 1.047& 2.597 & 0.955
&5.835 & 1.063 & 5.943\\
\end{tabular}
\end{table} 

\begin{table}
\caption{Mass spectrum of $D$ mesons with the account of $1/m_Q$
corrections in comparison with other quark model predictions and
experimental data. All masses are given in GeV. We use the notation
$(nL_J)$ for meson states, where $J$ is the total angular momentum
of the meson.}
\label{md}

\begin{tabular}{ccccc}
State& our & \cite{gi} & \cite{i}&  experiment \cite{PDG,orb,rad}\\
\hline
$1S_0$ & 1.875 & 1.88 &       & 1.8645(5) \\
$1S_1$ & 2.009 & 2.04 &       & 2.0067(5) \\
$1P_2$ & 2.459 & 2.50 & 2.460 & 2.4589(20) \\
$1P_1$ & 2.414 & 2.47 & 2.415 & 2.4222(18) \\
$1P_1$ & 2.501 & 2.46 & 2.585 &       \\
$1P_0$ & 2.438 & 2.40 & 2.565 &       \\
$2S_0$ & 2.579 & 2.58 &       &       \\
$2S_1$ & 2.629 & 2.64 &       & 2.637(9) ? \\
\end{tabular}
\end{table}

\begin{table}
\caption{Mass spectrum of $D_s$ mesons with the account of $1/m_Q$
corrections in comparison with other quark model predictions and
experimental data. All masses are given in GeV. The same notation as in
Table~\ref{md} is used for meson states.}
\label{mds}

\begin{tabular}{ccccc}
State& our & \cite{gi} & \cite{ehq} & experiment \cite{PDG,orb,rad} \\
\hline
$1S_0$ & 1.981 & 1.98   &       & 1.9685(6)\\
$1S_1$ & 2.111 & 2.13   &       & 2.1124(7)\\
$1P_2$ & 2.560 & 2.59   & 2.561 & 2.5735(17) \\
$1P_1$ & 2.515 & 2.56   & 2.526 & 2.535(4) \\
$1P_1$ & 2.569 & 2.55   &       &       \\
$1P_0$ & 2.508 & 2.48   &       &       \\
$2S_0$ & 2.670 & 2.67   &       &       \\
$2S_1$ & 2.716 & 2.73   &       &       \\
\end{tabular}
\end{table}

\begin{table}
\caption{Mass spectrum of $B$ mesons with the account of $1/m_Q$
corrections in comparison with other quark model predictions and
experimental data. All masses are given in GeV. The  same notation as in
Table~\ref{md} is used for meson states.}
\label{mb}

\begin{tabular}{cccccc}
State & our & \cite{gi} & \cite{i} & \cite{ehq} & experiment
\cite{PDG,orb,rad}\\
\hline
$1S_0$ & 5.285 & 5.31 &       &      & 5.2792(18)\\
$1S_1$ & 5.324 & 5.37 &       &      & 5.3248(18) \\
$1P_2$ & 5.733 & 5.80 & 5.715 & 5.771& 5.730(9) \\
$1P_1$ & 5.719 & 5.78 & 5.700 & 5.759&        \\
$1P_1$ & 5.757 & 5.78 & 5.875 &      &      \\
$1P_0$ & 5.738 & 5.76 & 5.870 &      &        \\
$2S_0$ & 5.883 & 5.90 &       &      &       \\
$2S_1$ & 5.898 & 5.93 &       &      & 5.90(2) ? \\
\end{tabular}
\end{table}
 
\begin{table}
\caption{Mass spectrum of $B_s$ mesons with the account of $1/m_Q$
corrections in comparison with other quark model predictions and
experimental data. All masses are given in GeV. The same  notation as in
Table~\ref{md} is used for meson states.}
\label{mbs}

\begin{tabular}{ccccc}
State & our & \cite{gi} & \cite{ehq} & experiment \cite{PDG,orb,rad}\\
\hline
$1S_0$ & 5.375 & 5.39 &       & 5.3693(20)\\
$1S_1$ & 5.412 & 5.45 &       & 5.416(4) ? \\
$1P_2$ & 5.844 & 5.88 & 5.861 & 5.853(15) ? \\
$1P_1$ & 5.831 & 5.86 & 5.849 &         \\
$1P_1$ & 5.859 & 5.86 &       &         \\
$1P_0$ & 5.841 & 5.83 &       &          \\
$2S_0$ & 5.971 & 6.27 &       &          \\
$2S_1$ & 5.984 & 6.34 &       &          \\
\end{tabular}
\end{table}

\begin{table}
\caption{Hyperfine splittings of $P$ levels. All values are given in MeV.}
\label{splt}
\begin{tabular}{cccc|ccc}
States & $\Delta M_D$ & $\frac{m_c}{m_b}\Delta M_D$ & $\Delta M_B$ &
$\Delta M_{D_s}$ & $\frac{m_c}{m_b}\Delta M_{D_s}$ & $\Delta M_{B_s}$\\
\hline
$1P_2-1P_1$ & 45 & 14 & 14 & 45 & 14 & 13\\
$1P_1-1P_0$ & 63 & 20 & 19 & 61 & 19 & 18\\
\end{tabular}
\end{table}  

\begin{figure}[hbt]

\centerline{\begin{turn}{-90}\epsfxsize=13cm 
\epsfbox{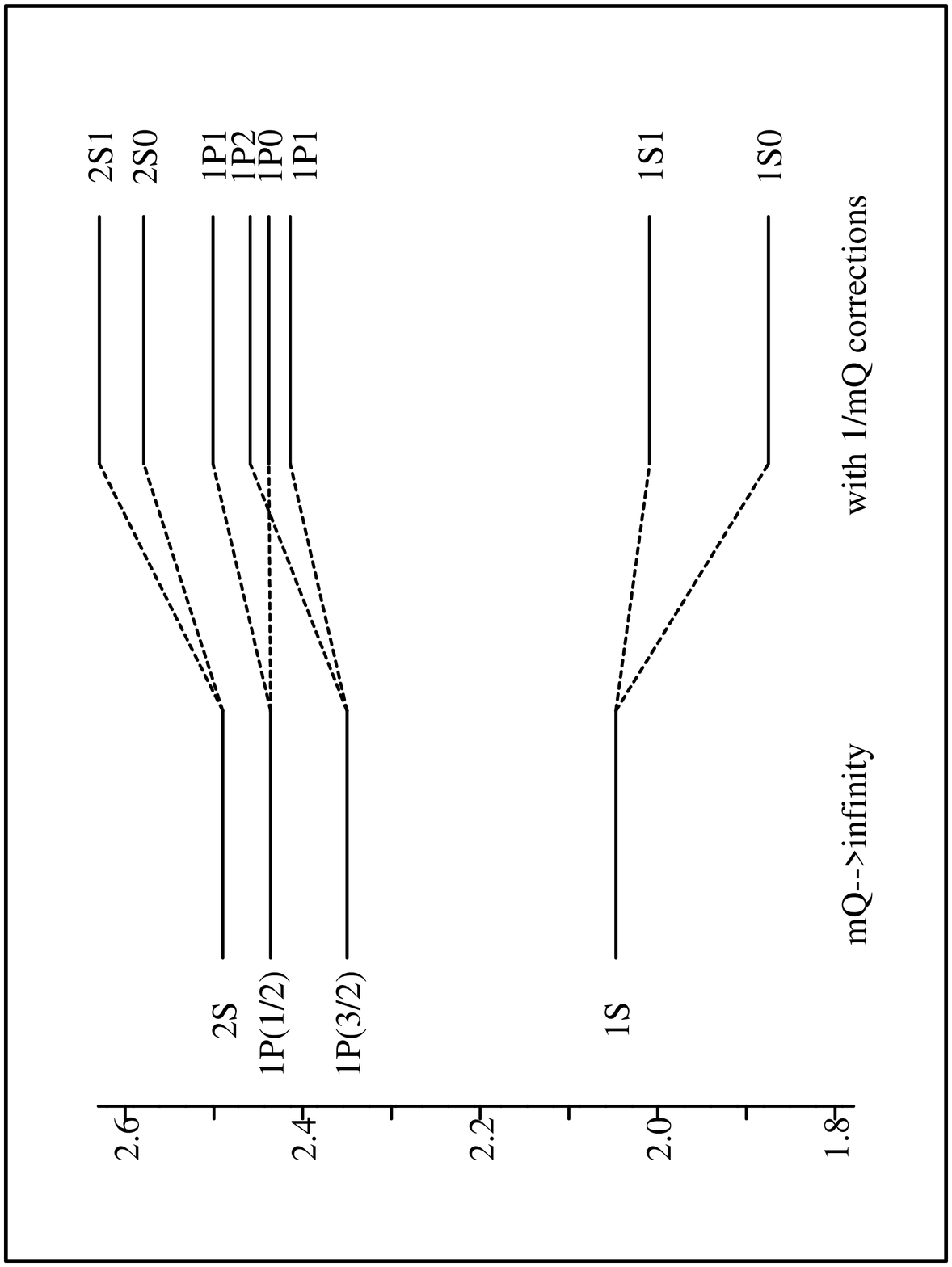}\end{turn}}

\caption{The ordering pattern of $D$ meson states. The mass scale is in GeV.}
\label{d}
\end{figure}

\begin{figure}[hbt]

\centerline{\begin{turn}{-90}\epsfxsize=13cm 
\epsfbox{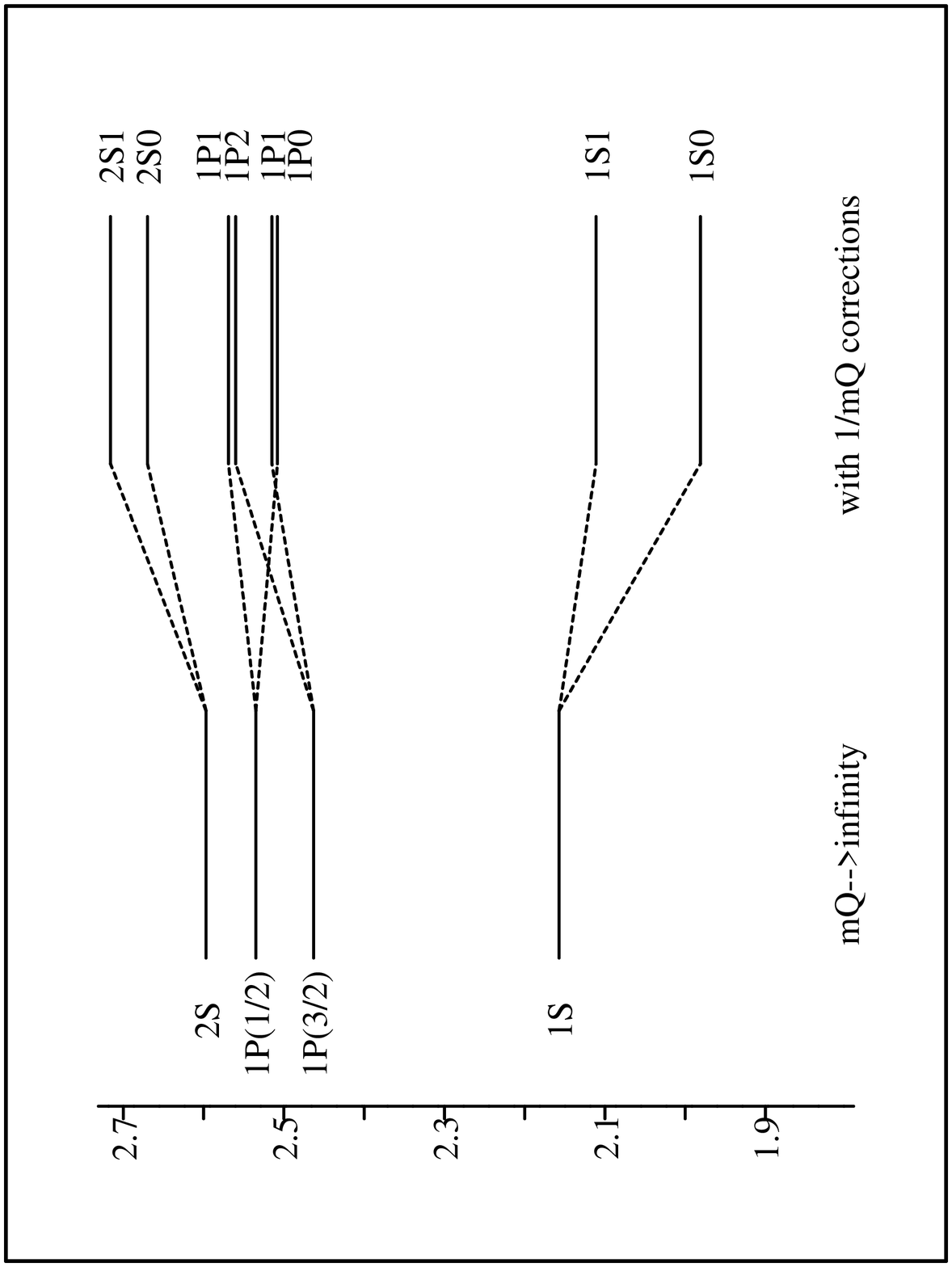}\end{turn}}

\caption{The ordering pattern of $D_s$ meson states. The mass scale is in GeV.}
\label{ds}
\end{figure}

\begin{figure}[hbt]

\centerline{\begin{turn}{-90}\epsfxsize=13cm 
\epsfbox{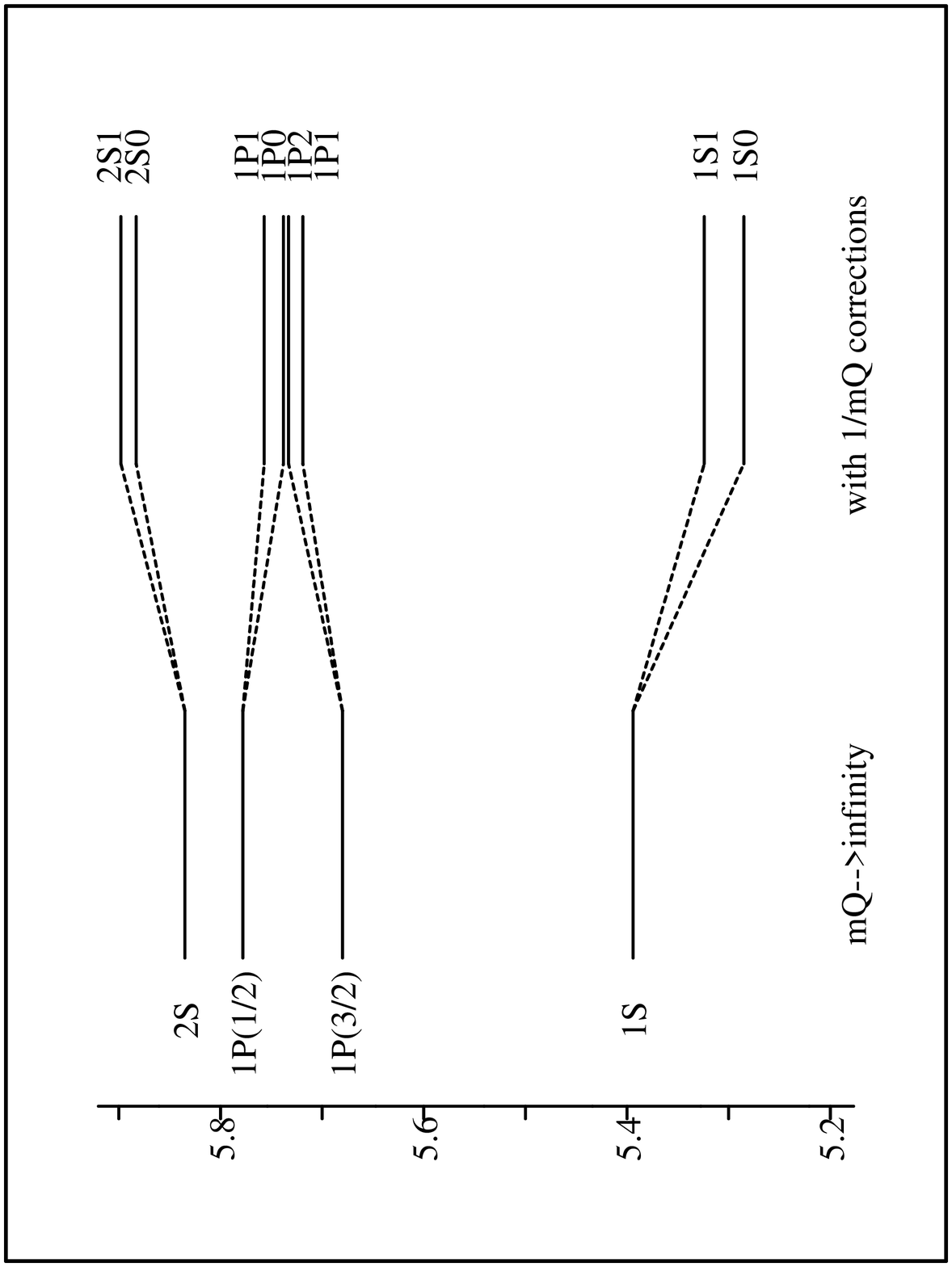}\end{turn}}

\caption{The ordering pattern of $B$ meson states. The mass scale is in GeV.}
\label{b}
\end{figure}

\begin{figure}[hbt]

\centerline{\begin{turn}{-90}\epsfxsize=13cm 
\epsfbox{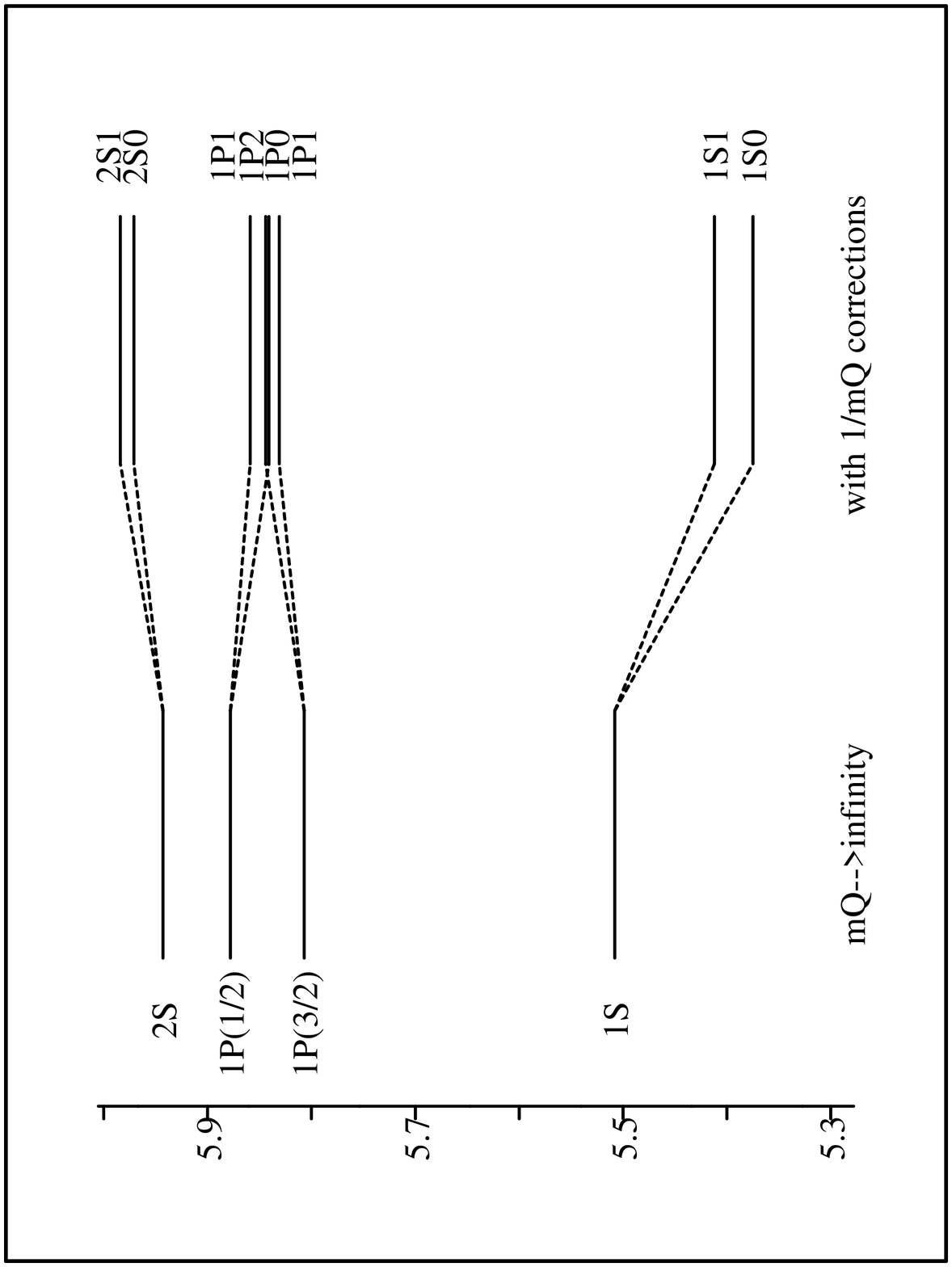}\end{turn}}

\caption{The ordering pattern of $B_s$ meson states. The mass scale is 
in GeV.}
\label{bs}
\end{figure}

\end{document}